\documentclass[twocolumn,amsmath,amssymb,aps,prb]{revtex4}
\usepackage{epsfig}
\usepackage{epstopdf}
\usepackage{graphicx}
\usepackage{grffile}
\usepackage{Jonasmacros}
\usepackage{dcolumn}
\usepackage{bm}
\usepackage[innercaption]{sidecap}
\sidecaptionvpos{figure}{t}
\usepackage{mathptmx}
\usepackage{txfonts}
\usepackage[colorlinks,citecolor=blue,linkcolor=blue]{hyperref}
\usepackage[usenames,dvipsnames,svgnames]{xcolor}
\usepackage{array}
\usepackage{float}
\usepackage{lipsum}
\usepackage{natbib}
\usepackage[normalem]{ulem}

\begin{document}

\renewcommand{\thefigure}{\arabic{figure}}
\def\be{\begin{align}}
\def\ee{\end{align}}
\def\ber{\begin{align}}
\def\eer{\end{align}}

\def\kv{{\bf k}}
\def\qv{{\bf q}}
\def\pv{{\bf p}}
\def\sigmav{{\bf \sigma}}
\def\tauv{{\bf \tau}}

\title{ Highly tunable magnetic coupling in ultrathin topological insulator films due to impurity resonances}

\author {Mahroo Shiranzaei}
\affiliation{Department of Physics and Astronomy, Uppsala University, Box 516, S-751 20 Uppsala, Sweden}
\author {Jonas Fransson}
\affiliation{Department of Physics and Astronomy, Uppsala University, Box 516, S-751 20 Uppsala, Sweden}
\author {Annica M. Black-Schaffer}
\affiliation{Department of Physics and Astronomy, Uppsala University, Box 516, S-751 20 Uppsala, Sweden}
\author {Fariborz Parhizgar}
\affiliation{Department of Physics and Astronomy, Uppsala University, Box 516, S-751 20 Uppsala, Sweden}

\begin{abstract}
We theoretically investigate the exchange interaction between magnetic impurities in ultrathin Bi$_2$Se$_3$ topological insulator films by taking into account the low-energy states produced by the impurities.
We find that the locally induced impurity resonances strongly influence the exchange interaction between magnetic moments. In particular, we find a non-collinear alignment being more favorable than the collinear ferromagnetic alignment preferred when impurity states are ignored and only the pristine topological insulator band structure is considered.
Moreover, we show that by applying of an electric field perpendicular to the ultrathin film, the exchange interaction can be drastically enhanced. This opens for the possibility of highly tunable magnetism by electric field.
\end{abstract}

\maketitle

\section{Introduction}
Magnetic impurity moments in solids couple indirectly to each other through the electronic carriers of the host material, with the nature of the coupling mechanism determined by the electronic structure of the host. In terms of the spatial extension, magnetic moments couple at distances up to a few nanometers in topological insulators, whereas the corresponding coupling in most semiconducting materials ranges only at most a few \AA ngstr\"{o}ms \cite{wray2011topological}. This long-range coupling between the magnetic impurities in topological insulators can lead to the existence of a robust magnetic phase, which has also been recently confirmed in experiments, both using angular resolved photoemission spectroscopy (ARPES) and scanning tunneling microscopy (STM) \cite{Zhang1582, PhysRevLett.106.206805, KOU201534}.

It has been suggested that the indirect magnetic coupling in topological insulators is governed by topology \cite{Zhang1582}. While the magnetically ordered phase in topologically trivial materials is fragile to decreasing dimensionality, the magnetic phase in topological insulators appears to be strengthened when lowering the dimension. Indeed, it has been observed that the magnetic phase generated by coupled magnetic impurities on the surface of topological insulators is far more robust than the corresponding phase created between bulk impurities \cite{Liu2088}. In line with this, magnetic coupling of impurities has also been verified in ultrathin films of topological insulators in experiments measuring the quantum anomalous Hall effect (QAHE) \cite{Chang167, chang2015high, checkelsky2014trajectory, PhysRevLett.113.137201, doi:10.1146/annurev-conmatphys-031115-011417}. Moreover, in topological insulators thinner than five quintuple layers, the surface states become gapped due to hybridization between the surface states at opposite sides \cite{NatPhys584}. Thus ultrathin films also provide an opportunity to combine the exotic properties of topological insulators and a finite energy gap into a single unit. In addition, with the band dispersion of ultrathin topological insulator films tunable by the application of an electric field, a tantalizing electric control of magnetism \cite{ohno2010window} might also be achievable in the ultrathin limit.

Very generally, the indirect magnetic coupling between magnetic impurities is governed by the magnetic susceptibility of the whole system. Thus, in materials with strong spin-orbit interaction, such as topological insulators, the magnetic susceptibility tensor gives three fundamentally different contributions: an isotropic (Heisenberg-like), a symmetrically anisotropic (Ising-like), and an asymmetrically anisotropic (Dzyalosinskii-Moryia-like) contribution \cite{PhysRevB.89.115431, PhysRevB.69.121303, PhysRevB.82.180411}. The isotropic and symmetrically anisotropic contributions lead to collinear configurations of the magnetic moments, whereas the asymmetrically anisotropic contribution favors non-collinear configurations \cite{PhysRevB.97.180402}. This asymmetric anisotropy easily leads to the existence of exotic magnetic phases, such as spin-glass, or chiral ferro- and anti-ferromagnetic phases \cite{PhysRevLett.106.136802, PhysRevB.94.045443}.

The nature of the indirect coupling between magnetic impurities is also significantly different in metallic and semiconducting materials. While the interaction in metals is dominated by the excitations around the Fermi level, or Ruderman-Kittel-Kausya-Yosida (RKKY) interactions, the interactions in semiconductors is mainly provided by the interband susceptibility between the valence and conduction band electrons, more related to van Vleck-type magnetism. The latter type based on interband susceptibility has in fact already been observed in topological insulator thin films \cite{yu2010quantized, PhysRevLett.114.146802, PhysRevLett.115.036805}.

With the indirect coupling determined by the magnetic susceptibility of the whole system, the impurities themselves can in fact modify the indirect coupling \cite{PhysRevB.68.235208, samir, PhysRevB.97.180402}. 
In particular, in topological insulators, impurity resonance states easily appear near the Dirac point due to scattering off the non-magnetic part of the impurity potential \cite{PhysRevB.81.233405, PhysRevB.85.121103, jonas-filling}. In ultrathin topological insulator films these resonances may even arise inside the band gap, depending on the strength of the impurity scattering potentials and the properties of the topological insulator \cite{PhysRevB.81.233405, PhysRevB.95.235429}, also recently addressed using \emph{ab-initio} calculations \cite{yu2010quantized, PhysRevB.97.155429, samir}. The gapped dispersion in ultrathin films thus makes impurity states even more prominent as they can be the only low-energy states available to mediate the indirect magnetic coupling. Despite the ubiquitousness of low-energy impurity states in topological insulators, most calculations of the magnetic susceptibility in topological insulator systems have been based only on the itinerant electrons of the unperturbed, or pristine, topological insulator surface states \cite{PhysRevLett.115.036805, PhysRevB.96.024413}, with only recent work highlighting the effect of the impurities, and then only in thick topological insulators \cite{PhysRevB.97.180402}. This is despite the fact that recent experiments on ultrathin topological insulator films have indicated that the indirect magnetic coupling depends strongly on the nature of the magnetic impurities \cite{Zhang1582, wray2011topological, Chang167, chang2015high, checkelsky2014trajectory, PhysRevLett.113.137201}.

In this work we therefore calculate the influence on the magnetic susceptibility of the impurity induced in-gap resonances in ultrathin topological insulator films, to accurately capture the indirect coupling between surface magnetic impurity moments. 
We show that the impurity resonances strongly enhance both the isotropic Heisenberg and symmetrically anisotropic Ising contributions. We also find a strong energy dependence, which  opens up for excellent tuning possibilities of the indirect coupling on- and off-resonance. Most importantly, we find that the asymmetrically anisotropic Dzyalosinskii-Moryia contribution is generated when we include the impurity states. As a direct consequence, a non-collinear collective configuration of the magnetic moments easily becomes favoured, with a finite out-of-plane net magnetization. This is in sharp contrast to the ferromagnetic ground state previously obtained when only considering the pristine ultrathin topological insulator films \cite{PhysRevLett.115.036805}. Finally, we also show that by simply applying an external electric field, the magnetic coupling becomes extensively tunable, between ferro- and anti-ferromagnetic to chiral configurations.

The remaining of the article is organized as follow. In Sec.~\ref{sec:model}, we introduce the model Hamiltonian and the general formalism for calculating the magnetic susceptibility, including the impurity states. In Sec.~\ref{sec:res} we present our results, with a focus on the contribution from impurity resonances and their tunability. Finally, we summarize and offer a few concluding remarks in Sec.~\ref{sec:con}. Some detailed part of the calculations can be found in the appendices~\ref{appA}-\ref{appC}.

\section{Model and method}
\label{sec:model}
The low-energy properties of the surface state electrons in the ultrathin topological insulator films can be described by an effective two-dimensional Hamiltonian near the $\Gamma-$point describing the two surfaces
\begin{align}
\label{Hamil}
\Hamil_0({\bf k})=&
   	\tau_z\otimes
	\Bigl[
		\hbar v_{F}(\mathbf{k}\times\hat{\bf z})\cdot\boldsymbol{\sigma}
		 +
		 V\sigma_0
	\Bigr]
	+
	\Delta \tau_x\otimes\sigma_0
	+ \mu
	.
\end{align}
Here, $\boldsymbol{\sigma}$, $\boldsymbol{\tau}$ are Pauli matrices in the spin and surface space, respectively, $\mathbf{k} = (k_x,k_y)$ denotes the two-dimensional wave vector for the surface electrons, and $v_F$ is the Fermi velocity. Moreover, $V$ denotes the potential difference between the two surfaces. This potential arises from the combined effect of a substrate and/or an external electric field applied perpendicular to the film. Due to the thikness of the film there is an effective mass hybridization term $\Delta$ that couples the two surfaces of the topological insulator. The model in Eq.~(\ref{Hamil}) leads to the dispersion relation
\begin{align}
E_{sm}(k)=&
	\sign(s)\sqrt{\Bigl(\hbar v_{F} |\mathbf{k}|+(-1)^mV\Bigr)^2+\Delta^2},
\end{align}
of a gapped Dirac spectrum with gap size $2\Delta$. Here, $s = \pm 1$, refers to the conduction and the valence bands, respectively, whereas $ m = 1,2 $ labels the solutions.
Further, we model the added magnetic impurities through the Hamiltonian
\begin{align}
\label{Himp}
\Hamil=&
	\sum_i
	\Bigl(
		U\sigma_0
		+
		J_c\bfS_i\cdot\bfsigma
	\Bigr)\delta(\bfr-\bfr_i)
	,
\end{align}
where $U$ and $J_c\bfS_i$ represent the spin-independent and spin-dependent part, respectively, of the scattering potential. Here we allow magnetic impurities on sites $\bfr_i$, summing over all impurities, where we restrict ourselves to consider impurities in the top surface only, without limiting the applicability of our results. The only assumption in Eq.~\eqref{Himp} is that the impurities behave as classical spins with strength $|\bfS_i|=S$, as appropriate for higher spin impurities.

For two local impurity magnetic moments, located at $\bfr$ and $\bfr'$, respectively, the effective indirect coupling, or exchange, Hamiltonian can be written as
\begin{align}
\label{HRKKY}
\Hamil^\text{ex}=&
	\frac{1}{2}
		J_c^2
		\bfS_1
		\cdot
		\bfchi(\bfr,\bfr')
		\cdot
		\bfS_2
	,
\end{align}
where $\bfchi(\bfr,\bfr^\prime)$ is the magnetic susceptibility tensor.
Using the results in Ref.~\citep{PhysRevB.97.180402}, we can always write $\bfchi(\bfr,\bfr^\prime)$ as
\begin{align}
\label{spinsuscep}
\bfchi(\bfr,\bfr')=&
	\tr
	\int
		\frac{\bfsigma\boldsymbol{\calG}_0(\bfr,\bfr^\prime;\omega)\bfsigma \boldsymbol{\calG}_0(\bfr^\prime,\bfr;\omega)}
			{(1-Ug)^2-J_c^2S^2g^2}
	\frac{d\omega}{2\pi}
	.
\end{align}
In this expression, $\boldsymbol{\calG}_0(\bfr,\bfr';\omega)$ denotes the bare single electron Green's function, i.e.~without impurities, the trace runs over the spin degrees of freedom, $g(\omega)=\tr\int\boldsymbol{\calG}_0(\bfk,\omega)\;d\bfk/2$, and where the expression in the denominator encode for the contributions from the impurity states. The magnetic susceptibility of the pristine topological insulator, $\chi_0$, is simply retained by setting $U=J_cS=0$ in the denominator. We refer to Appendix \ref{appA} for more details on the bare Green's function.

It is convenient to rotate the spin vectors $\bfS_{i=1,2}$ into $\tilde{\bfS}_{i}=(S_{{i}x}\cos\varphi_R,S_{{i}y}\sin\varphi_R,S_{{i}z})$ \cite{PhysRevB.96.024413, PhysRevB.97.180402} in terms of the polar angle $\varphi_R$ of the relative distance between the impurities. Then the exchange Hamiltonian takes the form 
\begin{align}
\label{eq:newRKKY}
\Hamil^{\text{ex}}=&
	\biggl[
		H \; \bfS_1\cdot\bfS_2
		+
		I \; 
		\Big(
			\tilde{\bfS}_1\cdot\tilde{\bfS}_2
			+
			\tilde{S}_{1x}\tilde{S}_{2y}
			+
			\tilde{S}_{1y}\tilde{S}_{2x}
		\Big)
\nonumber\\&
		+
		\bfD
		\cdot
		\Bigl(
			\tilde{\bfS}_1\times\tilde{\bfS}_2
		\Bigr)
	\biggr]
	.
\end{align}
Here $H$ and $I$ refer to the isotropical and symmetrical anisotropic couplings, respectively, whereas $\bfD = D \;(1,-1,0)$ denotes the asymmetrical anisotropy. Here $H$, $I$, and ${\bf D}$ can be thought of as Heisenberg-, Ising-, and Dzyaloshinskii-Moriya-like interactions, respectively. Using the equation for the magnetic susceptibility given in Eq.~\eqref{spinsuscep} in Eq.~\eqref{HRKKY}, these interaction parameters are obtained from the expressions \cite{PhysRevB.97.180402}
\begin{subequations}
\label{eq:nonJs}
\begin{align}
\label{eq:nonJ}
H=&J_c^2
	\int
		\frac{\calG_{tt}^2(\bfr,\bfr^{\prime};\omega)+\calG_{tt}^{\prime2}(\bfr,\bfr^{\prime};\omega)}
			{(1-Ug)^2-J_c^2S^2g^2}
	\frac{d\omega}{\pi}
	,
\\
\label{eq:nonI}
I=&
	-2 J_c^2\int
		\frac{\calG_{tt}^{\prime2}(\bfr,\bfr^{\prime};\omega)}
			{(1-Ug)^2-J_c^2S^2g^2}
	\frac{d\omega}{\pi}
	,
\\
\label{eq:nonD}
D=&
	-2 J_c^2\int
		\frac{\calG_{tt}(\bfr,\bfr^{\prime};\omega)\calG_{tt}^{\prime}(\bfr,\bfr^{\prime};\omega)}
			{(1-Ug)^2-J_c^2S^2g^2}
	\frac{d\omega}{\pi}
	.
\end{align}
\end{subequations}
Here $\calG_{tt}$ and $\calG_{tt}^\prime$ indicate the $\uparrow\uparrow$ and $\uparrow\downarrow$ components, respectively, of the bare Green's function on the top surface, see Appendix \ref{appA} for more details. Again, the results for a pristine topological insulator, $J_0=H_0,I_0,D_0$, is obtained by setting $J_cS=U=0$ in the denominator of Eqs.~\eqref{eq:nonJs}.
In order to provide a physical understanding for the behavior of the exchange parameters in the presence of impurities, we also consider the spin-polarized local density of states (spin-resolved LDOS) $\rho_{\uparrow, \downarrow}$, see Appendix \ref{appB} for a detailed expression derived from the full Green's function.

Before proceeding we also note that the magnetic susceptibility is, in general, a sum of interband and intraband contributions, $\bfchi=\bfchi_\text{intra}+\bfchi_\text{inter}$. At zero temperature and finite occupancy at the Fermi level, such as in a metal, the intraband contribution normally dominates the exchange interactions. Then, the indirect magnetic coupling can be well described in terms of the itinerant electrons of the Fermi surface, giving rise to a (generalized) RKKY interaction. By contrast, in gapped systems, such as in pristine utrathin topological insulator films, the intraband contribution vanishes whenever the chemical potential lies within the band gap, due to the absence of carriers. Hence, only interband contributions are present, although they are also small \cite{PhysRevB.96.024413}. The part of the response function which originates from such an interband process between conduction and valence bands is usually referred to as a van Vleck interaction. \cite{yu2010quantized}
%%%%%%%%%%%%%%%%%%%%%%%%%%%%%%%%%%%%%%%%%%%%%%%%%
\section{Results}
\label{sec:res}
Using Eqs.~\eqref{eq:nonJs} we quantify the complete and general indirect exchange coupling in terms of $J = H, I$, and $D$ between two magnetic impurities on the top surface of an ultrathin topological insulator film. 
We first present the different components in an unbiased ($V=0$) film, and thereafter we discuss the effects of an external electric field. Building on these results, we proceed to study the magnetic ordering of impurity moments and the different magnetic phases in the system.

For the remainder of this work we assume a four quintuple layer thin film of Bi$_2$Se$_3$, which has an energy gap $2\Delta=70$ meV and Fermi velocity $v_F=4.48\times10^5$ m/s. We also assume that the chemical potential resides within the band gap, i.e.~$|\mu|<\Delta$, in order to directly connect which recent experiments \cite{doi:10.1146/annurev-conmatphys-031115-011417}.
Moreover, we assume an inter-impurity distance of $R=10$ \AA{} in the surface plane, unless we explicitly investigate the distance behavior. We also require that $0\leq J_cS\leq1$ and $0\leq U\leq6$ (both given in units of $\hbar v_F/k_c$, where $k_c$ is the band momentum cutoff) and keep $J_cS/U\approx1/8$. We find these values by comparing our spin-resolved LDOS with the results of Refs.~\cite{PhysRevLett.108.256811, Eelbo_2013, PhysRevB.89.104424, PhysRevB.96.235444}.
To simplify our plots we express the coupling terms, $J$, and spin-resolved LDOS in units of $(J_c/\hbar^2 v_F^2 \Omega_{BZ})^2$ and $1/\hbar^2 v_F^2 \Omega_{BZ}$, respectively, where $\Omega_{BZ}$ is the area of the first Brillouin zone.
\subsection{Impurity states and their exchange interaction contributions}
\begin{figure}[t]
\begin{center}
\includegraphics[width=\columnwidth]{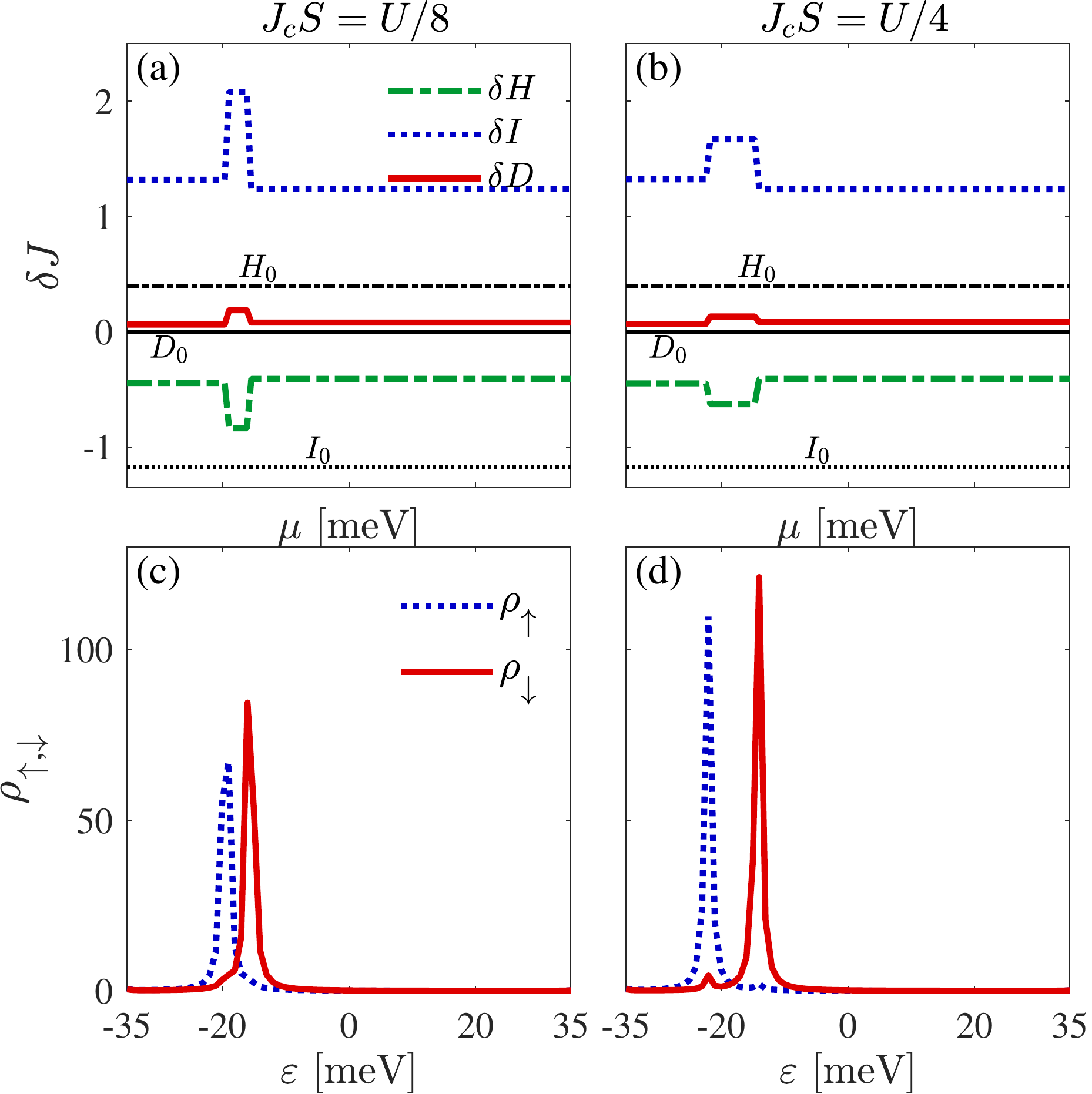}
\end{center}
\caption{(Color online) Exchange interaction $J = H, I, D$ contributions from the impurity states, $\delta J = J-J_0$, as a function of chemical potential $\mu$ for two impurities located at distance $R=10 \AA$ and for $V=0$, $U=4$ $\hbar v_F/k_c$, with $J_c S=U/8$ (a) and $U/4$ (b). Response of the pristine ultrathin topological insulator film is illustrated with black lines for comparison. (c,d) Corresponding spin-resolved LDOS  with respect to the energy $\varepsilon$ the Fermi level for one impurity at the location of the other impurity.}
\label{u4R10-new}
\end{figure}

We start by setting the parameters $V=0$ and $U=4$, in order to study how the impurities influence the exchange interaction. The plots in Fig.~\ref{u4R10-new}(a,b) show the corrections from the impurities, $\delta J=J-J_0$ with $J=H,\ D,\ I$ and $J_0=H_0,\ I_0,\ D_0$, as a function of the chemical potential $\mu$ within the energy gap for $J_cS=U/8$ and $J_cS=U/4$. As a reference we also plot the exchange interactions parameters for the pristine topological insulator, $J_0$, with black lines. 
In Fig.~\ref{u4R10-new}(c,d) we show the corresponding spin-resolved LDOS as a function of energy $\varepsilon$ at the Fermi level for the same values of $J_c$ and chemical potential $\mu = 0$ as produced by one impurity measured at the position of the other impurity. 

First, we directly see that the magnetic coupling is independent on the chemical potential, within the gap, in the pristine case, which is expected since this coupling is of van Vleck nature and thus given by interband transitions only. 
These constant exchange interactions obtained for the impurity-free ultrathin films should be contrasted with the interactions in the presence of impurities, which acquire both significantly different values and a very strong energy dependence, here reflected in the variation as a function of the chemical potential. In fact,
first, we observe that the correction $\delta J$ to the exchange interaction changes the overall amplitude of the exchange coupling. From this observation we conjecture that the carrier density redistributed from the valence band into the impurity resonances has a substantial influence on the magnetic susceptibility.
This is expected since the exchange coupling depends on the density of occupied states, i.e.~it is a Fermi sea property, and thus it is natural that the presence of impurity resonances contributes to the overall amplitude for a wide range of chemical potentials. In fact, we see that the isotropic and symmetrically anisotropic corrections $\delta H$ and $\delta I$ even obtain opposite signs compared to $H_0$ and $I_0$. The asymmetric anisotropy has, on the other hand, significantly different behavior. Since the band dispersion of a pristine topological insulator film is electron-hole symmetric, the overlap between states in the conduction and valence bands, due to their different helicities, result in a vanishing asymmetric anisotropy, $D_0$, for all chemical potential values inside the gap. But, in the presence of the impurity resonances, the electron-hole symmetry is broken and the asymmetric anisotropy becomes finite.
 
Second, there is a very strong energy dependence in a specific range of the chemical potential. More specifically, whenever the chemical potential  $\mu$ is positioned between the impurity resonances (see e.g.~the range $-20\lesssim\mu\lesssim-15$~meV in Fig.~\ref{u4R10-new}(a,c)), the amplitudes of the magnetic coupling are strongly enhanced compared to when both impurity resonances lie on the same side of the chemical potential. This strong energy dependence can thus be directly traced back to the emergence of impurity resonances inside the band gap, as seen in Figs.~\ref{u4R10-new}(c,d). As a consequence of the finite $J_cS$, the impurity resonances are spin split into two spin-polarized resonances, merging into one non-spin-polarized resonance in the limit $J_cS\rightarrow 0$ \cite{PhysRevB.95.235429}.

\begin{figure}[t]
\begin{center}
\includegraphics[width=\columnwidth]{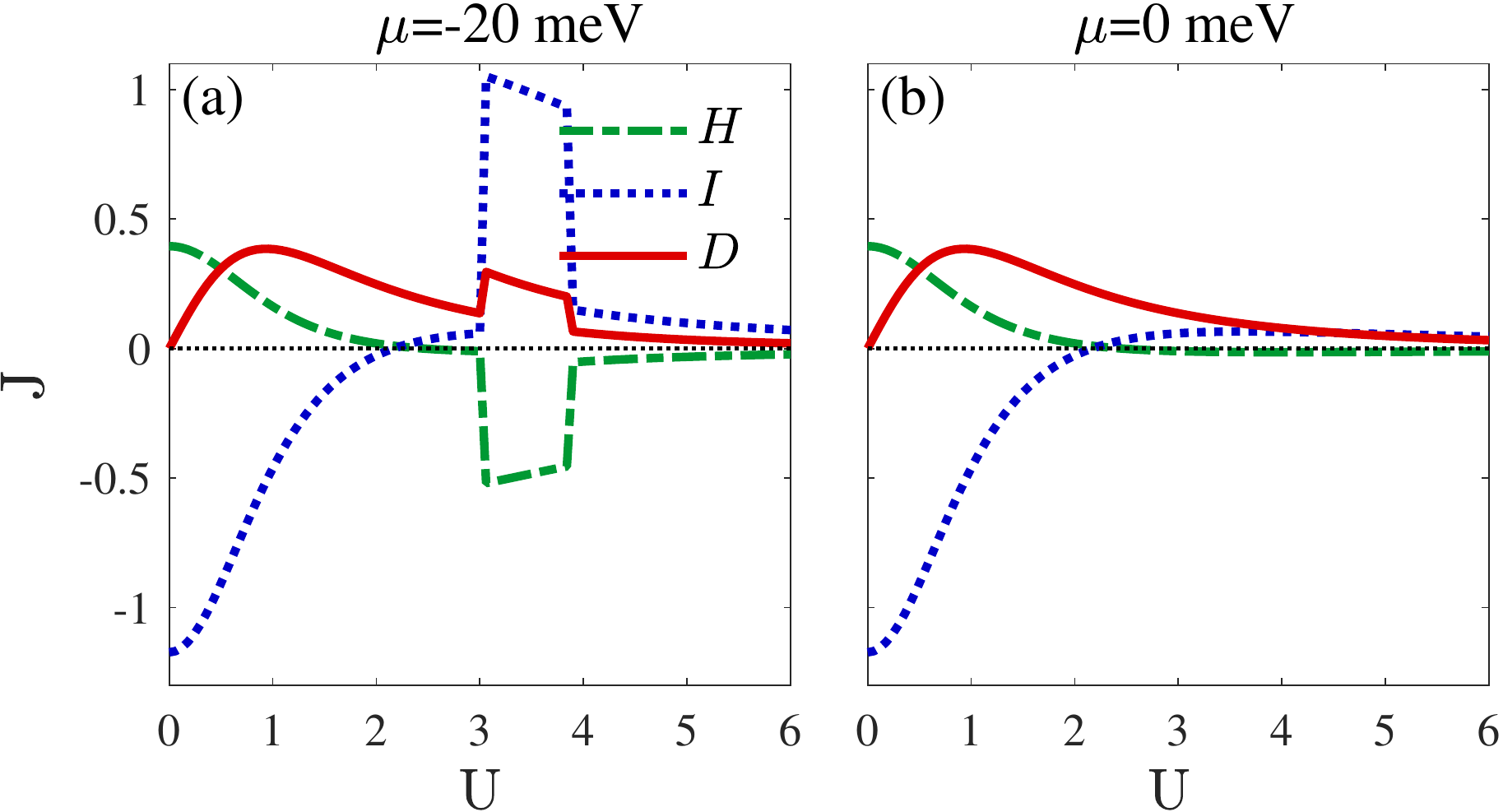}
\end{center}
\caption{(Color online) Exchange interactions $J=H, I, D$ as a function of $U$ (in units of $\hbar v_F/k_c$) for $R=10$ \AA, $V=0$ and $J_cS=U/8$ for $\mu= -20$ meV (a) and $\mu=0$ meV (b) and with dotted black lines marking zero.}
\label{JsumeFm002-new}
\end{figure}

Considering the properties of the impurities, both the spin-dependent $J_cS$ and spin-independent $U$ parts of the impurity potential can vary between the impurities. Therefore, in Fig.~\ref{JsumeFm002-new} we plot the exchange couplings $J=H,\ I,\ D$, as a function of the spin-independent impurity potential $U$, again keeping $R=10 \AA$  and $V=0$ and fixing $J_cS=U/8$, for two different chemical potentials (a) $\mu= -20$ meV and (b) $\mu=0$ meV. In both cases we observe that the isotropic (symmetric anisotropic) exchange is positive (negative) for low scattering potentials, but then transitions to negative (positive) values before diminishing for large $U$. The asymmetric anisotropy, on the other hand, vanishes in the absence of a scattering potential, peaks at small values before slowly approaching zero for increasing scattering potentials. However, in Fig.~\ref{JsumeFm002-new}(a) this overall smooth dependence on $U$ is interrupted by a sharply defined region with larger values, in the range $U\sim3$ to $U\sim4$. These boundaries exactly mark the energies where, at least one of, the induced impurity resonances coincide with $\mu$, in analogy with the sharp features in Fig.~\ref{u4R10-new}(a,b). The substantially increased values of the exchange coupling in this $U$-range thus originate from that the impurity resonances residing within the band gap. The absence of the corresponding features in Fig.~\ref{JsumeFm002-new}(b) is due to the fact that for this range of potentials $U$, the impurity resonances reside in the valence bands only and thus their effect is not present for the choice of in-gap value of $\mu$. 

\subsection{Electrical tunability of magnetism}
\begin{figure}[t]
\begin{center}
\includegraphics[width=\columnwidth]{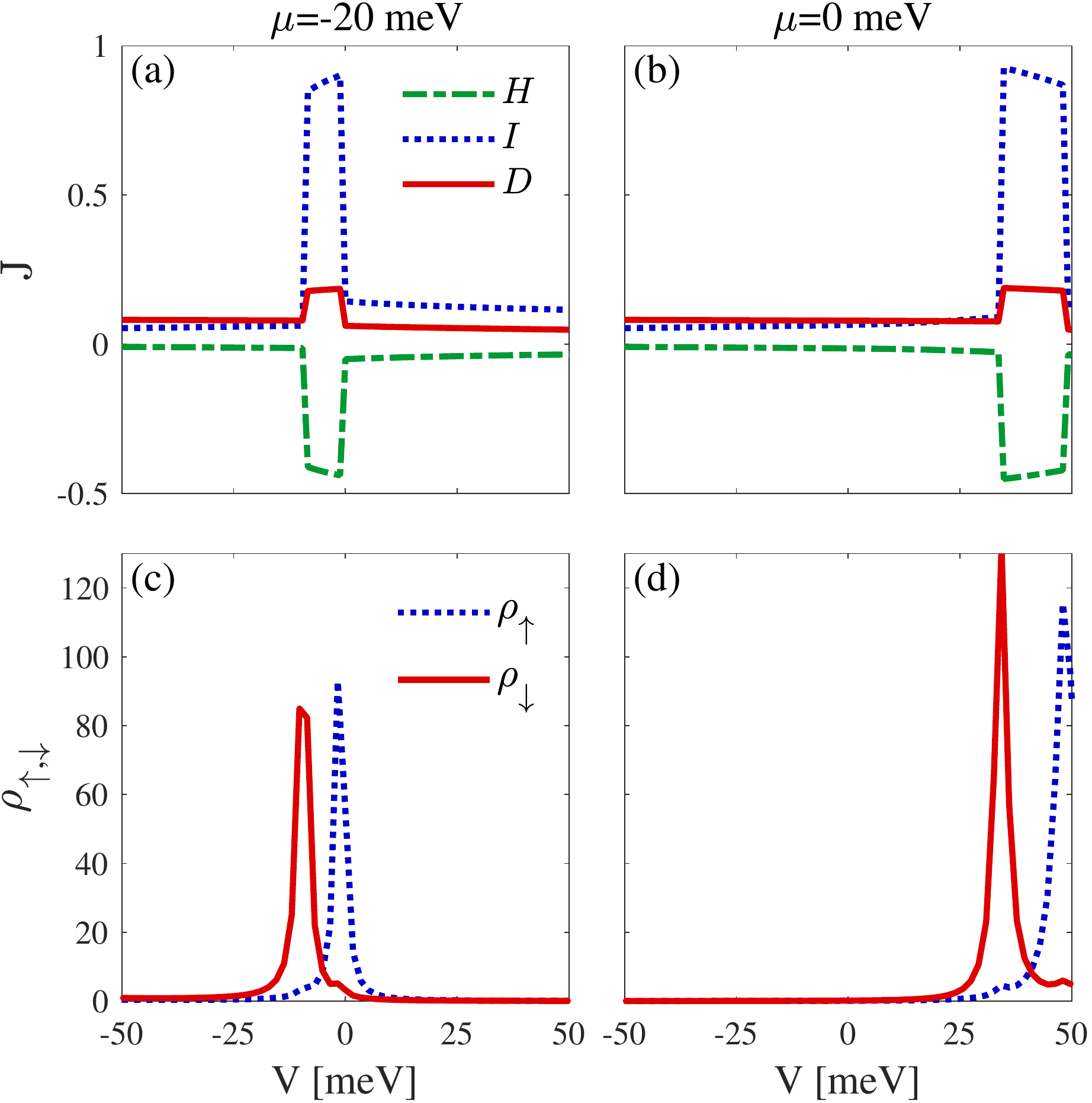}
\end{center}
\caption{(Color online) Exchange interactions $J=H, I, D$ as a function of $V$ for $U=4$, $J_c S=U/8$, $R=10 \; \AA$ and different chemical potentials $\mu=-20$ meV (a) and $\mu = 0$~meV (b). (c,d) Corresponding spin-resolved LDOS at energy $\varepsilon=\mu$ in for one impurity at the location of the other impurity.}
\label{Vu4m8R10-new}
\end{figure}
Having seen a strong dependence for the exchange coupling on the chemical potential, we next turn to the possibility to easily tune this behavior by applying an external electric field perpendicular to the plane of the film. 
To explore the signature of such potential difference in the magnetic exchange interaction, we present $H, I, D$ in terms of the parameter $V$ for different values of the chemical potential in Fig.~\ref{Vu4m8R10-new}(a,b), while in Fig.~\ref{Vu4m8R10-new}(c,d) we present the relevant spin-resolved LDOS extracted at energy $\varepsilon=\mu$ and plotted as a function of $V$. Clearly, we see how all the magnetic coupling terms are significantly higher for $V$s between two distinct values. In-between these two $V$-values the exchange coupling is in fact very large, for instance in (b) the Heisenberg coupling increases by $\sim 32$ times with respect to the unbiased case. For different chemical potentials the region of enhanced exchange coupling shifts, but it still exists equally prominently. The existence of a region with giant exchange couplings and its behavior with chemical potential and bias is explained by looking at the spin-resolved impurity resonance positions in Fig.~\ref{Vu4m8R10-new}(c,d). As been shown before in the Ref.~\cite{PhysRevB.95.235429}, the application of an electric potential alters the position of impurities resonance peaks inside the gap, which then also moves the corresponding region with giant exchange couplings. Thus we find a huge electric tunability, with an extreme sensitivity, of the exchange interactions in an ultrathin topological insulator film. It is worth mentioning here that qualitatively, all features exposed in Figs.~\ref{u4R10-new}-\ref{Vu4m8R10-new} are also valid for other impurity distances.

\subsection{Orientation of magnetic moments}
\begin{figure}[b]
\includegraphics[width=1.1\columnwidth]{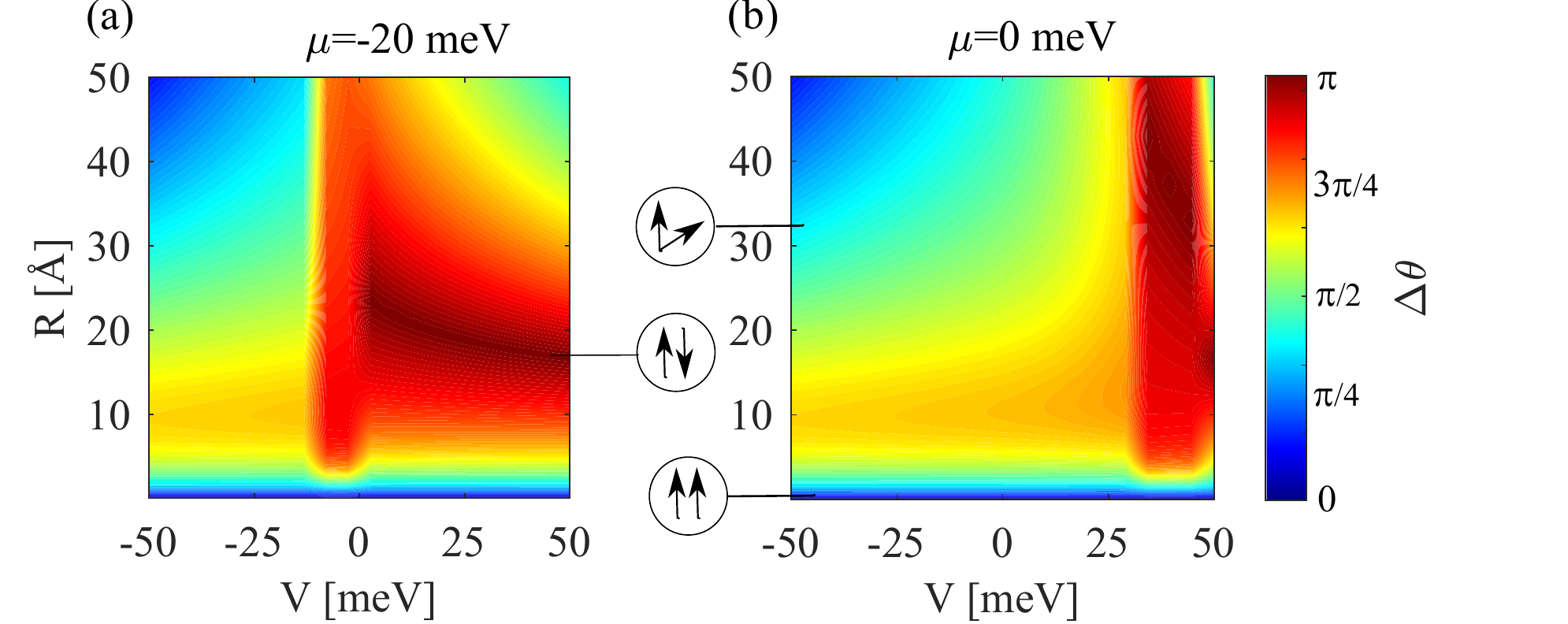}
\caption{(Color online) Contour-plots of energy-favourable angle between magnetic moments, $ \Delta\theta $, in the plane of $R-V$ for two different chemical potentials (a) $\mu=-20$ meV and (b) $\mu=0$. Here we set $ U=4$ $\hbar v_F/k_c$ and $J_cS = U/8$. The relative angles of moments is illustrated by arrows at some points.}
\label{phase}
\end{figure}

Having shown how two have a highly unusual magnetic impurities mutual interaction and also with a large tunability, we next calculate the spin configuration of two magnetic moments. We continue to assume classical spins, which means that the Hamiltonian \eqref{eq:newRKKY} can be rewritten as
\begin{align}\label{eq:order}
\Hamil^\text{ex}=&
	|S|^2 \; \Bigl[(H+I) \cos\theta_1\cos\theta_2
\nonumber\\&
	+
	H \;
	\Bigl(
		\cos\Delta\varphi
		+
		\cos\tilde{\varphi}_1\cos\tilde{\varphi}_2
	\Bigr)
	\sin\theta_1\sin\theta_2
\nonumber\\&
	+
	D \; \Bigl(\sin\theta_1\cos\theta_2\cos\tilde{\varphi}_1-\sin\theta_2\cos\theta_1\cos\tilde{\varphi}_2\Bigr)\Bigr],
\end{align}
where $\theta_{1,2}$ and $\varphi_{1,2}$ are the polar and azimuthal angles of the spin vectors, $\mathbf{S}_{1,2}$, respectively. Here, the azimuthal angles are considered with respect to $\varphi_R$, with $\tilde{\varphi}_{1(2)}=\varphi_{1(2)}-\varphi_R$ and we also define $\Delta\varphi=\varphi_2-\varphi_1$.
Following straightforward calculations presented in Appendix \ref{appC}, we find that the minimum energy of two magnetic impurities coupled to each other is given by $\tan\Delta\theta=D/(H+J)$, where $\Delta\theta=\theta_2-\theta_1$ and $\varphi_1=\varphi_2=\varphi_R$. The misalignment between the impurities is thus described in terms of the phase $\Delta\theta$. This phase is finite whenever the asymmetric anisotropy is finite ($D\neq0$), but vanishes in its absence.
By introducing new spin variables $\bar{S}(\theta)$ with $|\bar{S}|=|\bf S|$, azimuthal angle $\varphi=\varphi_R$, and polar angle $\theta$, the effective spin Hamiltonian can for this arrangement be written as $H=\bar{S}(0)\cdot \bar{S}(\Delta\theta)$ \cite{PhysRevB.69.121303}.

To further investigate the spin configurations, we plot in Fig.~\ref{phase} the relative polar angle $\Delta\theta$ between two magnetic moments, as a function of both the inter-impurity distance $R$ and electric field $V$ for both $\mu=-20$ meV (a) and $\mu=0$ (b). In both cases the configuration tends towards becoming ferromagnetic, i.e.~$\Delta\theta=0$, when the distance between the two impurities diminishes, $R\rightarrow 0$, for all values of $V$. At finite distances, however, the phase diagrams display a wide range of different configurations, spanning from ferromagnetic through non-collinear to anti-ferromagnetic configurations.
In particular, at both chemical potentials distinctive regimes exist where the moments align toward an anti-ferromagnetic-like configuration, $\Delta\theta > \pi/2$, for all distances $R>10$~\AA: in (a) $-10<V<0$~meV and in (b) $35<V<50$~meV. We trace these regimes directly trace back to the entry and exit of the chemical potential between the impurity resonances, see e.g.~Fig.~\ref{Vu4m8R10-new}(a,b). Beyond this electric field regime we find that most of the phase space is consists of clearly non-collinear configurations, where the relative angle is both far from 0 and $\pi$, with the exact configuration determined by $R$. This is due to the large influence from the asymmetric anisotropy, which renders the collinear cases less favorable compared to a non-collinear arrangement. 

Extending the results of Fig.~\ref{phase} to a multi-impurity setup, we conclude that impurities favour pair configurations that can be represented by the angle $\Delta\theta$ in the $\rho z$-plane, where $\hat{\rho}$ defines the in-plane direction between impurities. Since the impurities are located in different directions, $\hat{\rho}$, with respect to each other, the only common axis of all magnetic moments is along the $z$-axis. Hence, the resulting phase may most likely be ascribed a non-collinear ferromagnetic nature, but with generally a $z$-axis component, i.e.~an out-of-plane common component. 

\section{Concluding remarks}\label{sec:con}
In summary, we have investigated effects of impurity resonances on the magnetic exchange coupling between the magnetic moments located on the surface of ultrathin topological insulator films. We find that the contribution from the impurity resonances, typically, is of the same order, as the bare contribution originating from the unperturbed surface states, but become much larger under certain resonance conditions. 
We further analyze the importance of the impurity resonances on the magnetic interactions in terms of the isotropic, and symmetric and anti-symmetric anisotropy components. For a pristine surface, the first two components are finite whereas the last one vanishes identically. 
We find that the contribution from the impurity states on the symmetric anisotropy and isotropic components, which both leads to collinear alignment of the magnetic moments moments, are within the same order of magnitude as the corresponding contributions from the intrinsic electronic structure, however, with opposite signs. Overall, this has a tendency to lead to a weakened collinear coupling. Most importantly, the non-collinear asymmetric anisotropic interaction, which is zero for pristine films, acquires a large contribution from the impurity states and imply that the collective ground state
of the magnetic impurities should be strongly non-collinear.

Furthermore, we show that the applications of an electric field perpendicular to the ultrathin film can be used as a mechanism to shift the energy of the impurity resonances, which opens up the possibility to electrically tune the properties of the magnetic interactions. In effect, this mechanism should provide a tool for tuning between isotropic and anisotropic interactions, something which clearly has a great impact on the magnetic state.

Based on our findings we conclude that calculations of the magnetic exchange interactions, without considering effects originating from the impurities themselves are over-simplified \cite{PhysRevLett.106.097201}. Hence, there is a great risk of losing interesting and important features in the system. In particular, the exchange interaction based on the pristine topological insulator surface states misses the anti-symmetric anisotropic component, which inevitable leads to the prediction of a non-collinear ferromagnetic phase. In fact, the easy-axis of many magnetic impurities on the surface of topological insulators is in-plane \cite{PhysRevB.97.155429}, where calculations using the pristine system give in-plane ferromagnetic phase, while some experiments have already shown a perpendicular magnetic phase \cite{Chen659, chang2013thin}. 
However, taking into account the influence from the impurity states but merely including the isotropic and symmetric anisotropic components of the exchange coupling $(H+I)$ does not provide a sufficient description, as it leads to an anti-ferromagnetic phase. Here, we have shown the importance of the anti-symmetric anisotropy and the necessity to include it in calculations of the magnetic phase of ultrathin topological insulators films. It is this latter component term that leads to a chiral ferromagnetic phase.
Such chiral magnetic phase will clearly affect previous theoretical studies on QAHE experiments \cite{doi:10.1146/annurev-conmatphys-033117-054144}. The QAHE has previously been assumed to be proportional to the net magnetization in the system. However, it has more recently been shown that the QAHE persists also in chiral ferromagnet and chiral antiferromagnetic systems with zero net magnetization \cite{PhysRevLett.112.017205}, extending the effect to large parts of the phase diagram uncovered in this work.

\section{Acknowledgment}
M.~Sh.~and J.~F.~thank the Carl Tryggers Stiftelse and Vetenskapsr\aa det for financial support.
A.~B.-S.~and F.~P.~acknowledge financial support from the Swedish Research Council (Vetenskapsr\aa det Grant No.~2018-03488) and the Knut and Alice Wallenberg Foundation through the Wallenberg Academy Fellows program. M.~Sh.~also thanks the Institute for Research in Fundamental Sciences (IPM) for their hospitality.

\begin{widetext}
\appendix 

\section{Real space Green's function}\label{appA}
In this Appendix we provide the components of the bare Green's function, i.e.~for the pristine ultrathin topological insulator film without any impurities, which is used in the main text in Eq.~\eqref{spinsuscep} and also the on-site bare Green's function $g(\omega)$. 
The Matsubara Green's function in reciprocal space is given by $\mathcal{G}_0({\mathbf{k}};\omega)=[i\omega+\mu-H_0]^{-1}$, which can be transformed into real-space by taking the Fourier transformation as 
\begin{equation}\label{eq:appA_1}
\begin{split}
 \mathcal{G}_0(R;\omega)=\frac{1}{\Omega_{BZ}}\int \ d\mathbf{k} \ e^{i \, \mathbf{k} \cdot \mathbf{R}} \, \mathcal{G}_0({\mathbf{k}};\omega).
      \end{split}
\end{equation}
After some straightforward calculations, the general matrix form of the bare Green's function is a $4 \times 4$ matrix reading
\begin{align}\label{eq:appA_2}
\mathcal{G}_0(\pm R;\omega)=
\begin{bmatrix}
    \mathcal{G}_{tt} & \mp e^{-i \, \varphi_R} \, \mathcal{G}^\prime_{tt}  & \mathcal{G}_{tb} & \mp e^{-i \, \varphi_R} \, \mathcal{G}^\prime_{tb} \\
   \pm e^{i \, \varphi_R} \, \mathcal{G}^\prime_{tt} & \mathcal{G}_{tt}  & \pm e^{i \, \varphi_R} \, \mathcal{G}^\prime_{tb} & \mathcal{G}_{tb} \\
     \mathcal{G}_{tb} & \mp e^{-i \, \varphi_R} \, \mathcal{G}^\prime_{tb} &  \mathcal{G}_{bb} & \mp e^{-i \, \varphi_R} \, \mathcal{G}^\prime_{bb} \\
    \pm e^{i \, \varphi_R} \, \mathcal{G}^\prime_{tb} & \mathcal{G}_{tb}  & \pm e^{i \, \varphi_R} \, \mathcal{G}^\prime_{bb}& G_{bb} \\
  \end{bmatrix},
\end{align}
where $ \varphi_R = \arctan \: (R_y/R_x) $ denotes the polar angle of the vector $ {\bf R}=\bfr-\bfr' $ between the two impurities. As we assume the impurities to be located on the top surface of the ultrathin topological insulator film, we need the upper right block of this matrix, in which $\mathcal{G}_{tt}$ and $\mathcal{G}^\prime_{tt}$ represent the $\uparrow \uparrow$ and $\uparrow \downarrow$ spin configurations of the bare Green's function, respectively, given by
\begin{subequations}
\begin{align}
\mathcal{G}_{tt} (R; \omega)=&
	\frac{\pi}{\hbar^2 v_F^2 \Omega_{BZ}}
	\sum_{s=\pm}
	\biggl(1+\frac{s\bar{\omega}}{\sqrt{\bar{\omega}^2-\Delta^2}}\biggr) \;
	\biggl(V-s\sqrt{\bar{\omega}^2-\Delta^2}\biggr) \;
	K_0(\tilde{R})
\\ \nonumber \\ 
\mathcal{G}^\prime_{tt} (R; \omega)=&
	\frac{-i \; \pi}{\hbar^2 v_F^2 \Omega_{BZ}}
	\sum_{s=\pm}
	\biggl(1+s\frac{\bar{\omega}}{\sqrt{\bar{\omega}^2-\Delta^2}}\biggr) \;
	|V-s\bar{\omega}| \;
	K_1(\tilde{R})
\end{align}
\end{subequations}
where $\bar{\omega}=i \omega + \mu$ and $\tilde{R} = i \,  R \, ( V - s \sqrt{\bar{\omega}^2-\Delta^2} ) \, / \, \hbar v_F$, and $K_{0,1}$ are the modified Bessel functions of the second kind. 
From these expressions we also find the analytic expression of the on-site Green's function
\begin{equation}
\begin{aligned}
g = \frac{\pi}{2 \; \hbar^2 v^2_F \Omega_{BZ}} \Bigg[ & (V-\bar{\omega}) \ln  \frac{\Big( ( \hbar v_F k_c-V)^2-\bar{\omega}^2 +\Delta^2 \Big) \; \Big( (\hbar v_F k_c+V)^2-\bar{\omega}^2 +\Delta^2 \Big)}{\big( V^2+\Delta^2-\bar{\omega}^2 \big)^2}
\\ &
+2 \; \frac{V \bar{\omega} + \Delta^2 -\bar{\omega}^2}{\sqrt{\Delta^2 -\bar{\omega}^2}}  
 \arctan \frac{ 2 V^2  \big(\Delta^2 + \hbar^2 v^2_F k^2_c +\bar{\omega}^2- V^2 \big) }{\big(\Delta^2+\bar{\omega}^2-V^2 \big)^2 - V^4} 
\Bigg].
\end{aligned}
\end{equation}

\section{Spin-resolved LDOS}\label{appB}
In this Appendix we provide the expression for the spin-resolved LDOS of an ultrathin topological insulator film in the presence of a single impurity.
We use the T-matrix approach to find the dressed Green's functions. The T-matrix approach allows for a simple treatment of the scattering of surface Rashba-type electrons from a single impurity placed on the surface of the ultrathin topological insulator film. By considering a single impurity on the top surface including both an electrostatic and magnetic scattering potential as in Eq.~\eqref{Himp}, the LDOS for spin up and down electrons on the top surface of the total system can be obtained from the expression
\begin{align}
\boldsymbol{\rho}_\pm=-\frac{1}{2\pi} \text{Im} \;\; \text{tr} \Big[ (\sigma_0 \pm \boldsymbol{\sigma}) \, G(R;\varepsilon) \Big],
\end{align}
where $G(R;\varepsilon)$ refers to retarded Green's function, obtained from the Matsubara Green's function $\mathcal{G}(R;\omega)$, by letting $i \omega \rightarrow \varepsilon + i 0^+$. For a $\hat{z}$-axis polarized magnetic impurity, the spin-resolved LDOS is given by, see also Ref.~\cite{PhysRevB.96.024413},
%%%%%%%%%%
\begin{align}\label{eq:sldosup}
\begin{split}
\rho^{\uparrow} &=\frac{-1}{\pi} \text{Im} \Bigg[
g
+
\frac{4 \pi ^2 (J_c S + U)}{g (J_c S + U)-1} \Bigg(\sum_{s=\pm} s \, \big(V-i \, s \gamma \big)K_0(\tilde{R}) \Bigg)^2  
+
\frac{4 \pi ^2 (J_c S - U) }{g (J_c S - U)+1} \Bigg(\sum_{s=\pm} i a_s s (V-i \, s \gamma )K_1(\tilde{R})  \Bigg)^2 \Bigg],
   \end{split}
\end{align}
\begin{align}\label{eq:sldosdown}
\begin{split}
\rho^{\downarrow} &=\frac{-1}{\pi} \text{Im} \Bigg[
g
+
\frac{4 \pi ^2 (J_c S - U )}{g (J_c S - U)+1} \Bigg(\sum_{s=\pm} s \, \big(V-i \, s \gamma \big) K_0(\tilde{R}) \Bigg)^2
+
\frac{4 \pi ^2 (J_c S + U) }{g (J_c S + U)-1} \Bigg(\sum_{s=\pm} i a_s s (V-i \, s \gamma )K_1(\tilde{R})  \Bigg)^2 \Bigg].
   \end{split}
\end{align}
%%%%%%%%%%
In the above equations, $\gamma=\sqrt{\Delta ^2-(\varepsilon +i \, 0^+ )^2}$ and $a_s=\frac{1}{2} (\frac{\varepsilon +i \, 0^+ }{\gamma} + i  s )$, with the remaining quantities defined in the main text or in Appendix \ref{appA}.

\section{Spin-ordering}\label{appC}
In this Appendix we discuss in more detail how to find the minimizing condition for the relative spin configuration of two magnetic moments. Equation~\eqref{eq:order} in the main text is written in a rotated basis around the $\hat{z}$-direction by an angle $\varphi_R$, which makes the $\hat{x}$-axis along the direction between the two impurities. By applying another rotation of angle $-\pi/2$ around the newly defined $\hat{x}$-axis, the $\hat{y}$-axis is interchanged with $\hat{z}$: $\hat{z}\rightarrow -\hat{y}$. Then in this new basis, the Hamiltonian is written as
\begin{align}
\Hamil^{ex}=|\mathbf{S}|^2 \Big( H \cos \theta^\prime_1 \cos \theta^\prime_2 + (H+I) \cos (\varphi^\prime_1-\varphi_2^\prime) \sin \theta^\prime_1 \sin \theta^\prime_2 + D  \sin(\varphi^\prime_1-\varphi_2^\prime) \sin \theta^\prime_1 \sin \theta^\prime_2 \Big),
\end{align}
where $\theta^\prime$ and $\varphi^\prime$ are the polar and azimuthal angles, respectively, of the spin vectors in the new basis. The benefit of working in this new basis is that the Hamiltonian is dependent only to three angles as it is only related to the difference between azimuthal angles, $\phi=\varphi^\prime_1-\varphi^\prime_2$. By minimizing the Hamiltonian with respect to these three angles we find that the extrema of the system occur either at one of the five following points: $(\theta^\prime_1,\theta^\prime_2)=$\,$(0,0)$,$(0,\pi)$,$(\pi,0)$,$(\pi,\pi)$, or $(\pi/2,\pi/2)$. For the first four points the azimuthal angles are not well-defined and the Hessian matrix, which define the concavity of the system, is zero and hence the system is at a saddle point. 
At the last point, $(\theta^\prime_1,\theta^\prime_2)=(\pi/2,\pi/2)$, the minimization condition occurs for the relative azimuthal angle $\phi = \arctan (D/(H+I))$.
For a true minimum the determinant of the Hessian matrix, given by
\begin{align}
{\cal D}=-x(x^2-J^2)>0, \,\,\, x = (H+I) \cos \phi + D \sin \phi ,
\end{align}
should be positive \cite{PhysRevB.87.125401, adams2003calculus}.
With two $\phi$s satisfying the relation $\tan \phi =D/(H+I)$, we take the solution that makes ${\cal D}$ positive and arrive at the minimum energy configuration.
\end{widetext}

\bibliography{Magnetic_coupling_TI_thin_film}

\end{document}